\documentclass[10pt,twocolumn]{article}
\newcommand{\PRL}{Phys. Rev. Lett.}
\newcommand{\PR}{Phys. Rev.}
\newcommand{\PL}{Phys. Lett.}
\newcommand{\NIM}{Nucl. Instr. Meth.}
\usepackage{graphicx,epsfig} 
\def\etal{\emph{et al.}}

\begin{document}
 
\title{$\bar\Lambda$--hyperon global polarization in Au+Au collisions at RHIC}
\author{I Selyuzhenkov for the STAR Collaboration
\\ {\it \small Wayne State University, 666 W Hancock, Detroit MI 48201, USA}
\\ {\small {\it E-mail:} Ilya.Selyuzhenkov@wayne.edu}}
\date{}
\twocolumn[
\begin{@twocolumnfalse}
\maketitle
\begin{abstract} 
We present the $\bar\Lambda$--hyperon global polarization in Au+Au collisions
at $\sqrt{s_{NN}}=62$~GeV and $200$~GeV measured with the STAR detector at RHIC.
The observed $\bar\Lambda$--hyperon global polarization is consistent with zero,
what is in agreement with recent measurements of $\Lambda$ global polarization, as well as
$\phi(1020)$ and ${K^*}^0(892)$ vector mesons spin alignment with respect to the reaction plane.
The possible dependence of the global polarization on relative azimuthal angle
between the orbital momentum of the system and the hyperon 3-momentum is discussed.
The corresponding systematic uncertainty due to detector acceptance is found to be less than 20\%.
\\
\end{abstract} 
\end{@twocolumnfalse}
]

\section{Introduction}
\label{Introduction} 

Particles produced in non-central relativistic nucleus-nucleus collisions
are predicted to be globally polarized
along the direction of the system's orbital angular momentum,
perpendicular to the reaction plane~\cite{LiangPRL94, Voloshin0410089, Liang0411101}.
The origin of this global polarization is in the transformation
of the orbital angular momentum into the particle's spin due to spin-orbital coupling.
Among different observable consequences of this effect
are the hyperon's global polarization and global spin alignment of vector mesons.

The global spin-orbital transformation can happen at various evolution stages of the system
and its observation can provide important information on the hadronization mechanism and the origin of particle's spin.
One specific scenario for such spin-orbit transformation via the polarized quark phase is
discussed in~\cite{LiangPRL94}. Assuming that the strange and non-strange quark polarizations,
$P_s$ and $P_q$, are equal, in the particular case of
`exclusive' parton recombination scenario~\cite{LiangPRL94}, the values of the global
polarization $P_H$ for $\Lambda$, $\Sigma$, and $\Xi$ hyperons appear to be similar to
those for quarks: $P_H = P_q \simeq 0.3$.
At the same time, the predicted global spin alignment of vector mesons
is defined by terms proportional to higher quark polarization powers, $P_{q}^2$~\cite{ Liang0411101}.
Recently more realistic calculations~\cite{Liang:Xian2006}
of the global quark polarization were performed within a model based
on the HTL (Hard Thermal Loop) gluon propagator. The resulting hyperon polarization was
predicted to be in the range from $-0.03$ to $0.15$ depending on the temperature of the
QGP formed.

Preliminary results of $\Lambda$--hyperon global polarization, $\phi(1020)$ and ${K^*}^0(892)$
vector meson's spin alignment with respect to reaction plane
were recently presented~\cite{Selyuzhenkov:2005xa, Selyuzhenkov:2006fc, Chen:2007}.
In this paper we present the results for $\bar\Lambda$--hyperon global
polarization in Au+Au collisions   at $\sqrt{s_{NN}}$=62 and
200~GeV as a function of $\bar\Lambda$ transverse momentum and pseudorapidity
measured with the STAR (Solenoidal Tracker At RHIC) detector.

\section{$\bar\Lambda$ global polarization}
\label{Polarization}
$\bar\Lambda$ global polarization can be determined from the angular distribution of its decay
products with respect to the system orbital momentum {\boldmath $L$}:
\begin{eqnarray}
\label{GlobalPolarizationDefinition}
\frac{dN}{d \cos\theta^*} \sim 1~+~\alpha_{\bar\Lambda}~P_{\bar\Lambda}~\cos \theta^*,
\end{eqnarray}
where $P_{\bar\Lambda}$ is the $\bar\Lambda$ global polarization,
$\alpha_{\bar\Lambda} = - 0.642\pm0.013$~\cite{Eidelman:2004wy} is the $\bar\Lambda$ decay parameter,
$\theta^*$ is the angle between the system's orbital momentum {\boldmath $L$}
and the 3-momentum of $\bar\Lambda$'s decay products in the $\bar\Lambda$'s rest frame.

The observable used in the $\bar\Lambda$ global polarization measurement is derived in~\cite{Selyuzhenkov:2006fc}:
\begin{eqnarray} 
\label{GlobalPolarizationObservable} 
P_{\bar\Lambda}~=~\frac{8}{\pi\alpha_{\bar\Lambda}}\langle \sin \left( \phi^*_p 
- \Psi_{RP}\right)\rangle.
\end{eqnarray}
Here $\phi^*_p$ is the azimuthal angle of the anti-proton's 3-momentum, measured in $\bar\Lambda$'s rest frame.
Angle brackets in this equation denote averaging over the solid angle of anti-proton's 3-momentum
in $\bar\Lambda$'s rest frame and over all directions of the system 
orbital momentum {\boldmath $L$}, or, in other words, over all possible orientations of 
the reaction plane.

In this paper, $\bar\Lambda$ particles were reconstructed from their weak decay topology, $\bar\Lambda \to \bar p \pi^+ $,
using charged tracks measured in the STAR main TPC (Time Projection Chamber)~\cite{Anderson:2003ur}.
The reaction plane angle in Eq.~\ref{GlobalPolarizationObservable} is estimated by 
calculating the so-called event plane flow vector $Q_{EP}$~\cite{Voloshin:1994mz,Poskanzer:1998yz}.
This first-order event plane vector was determined from
charged tracks measured in two STAR Forward TPCs~\cite{Ackermann:2002yx}.
\section{Results}
\label{Results}
\begin{figure}[h]
\begin{center}
\includegraphics[width=0.5\textwidth]{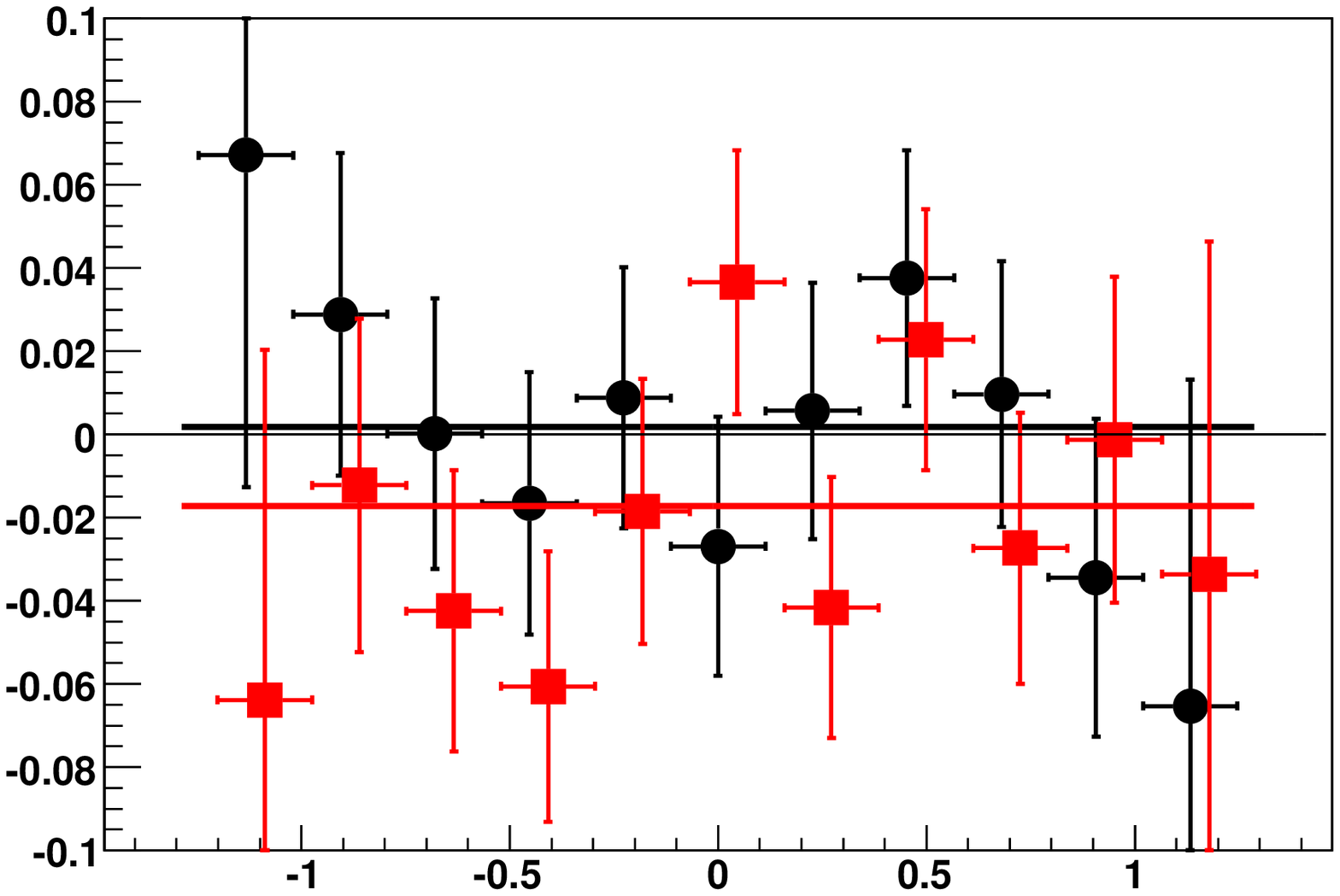}%
\put(-230,70){\rotatebox{90}{$P_{\bar\Lambda}$}} 
\put(-120,-5){$\eta$} 
\put(-128,125){{\bf STAR Preliminary}}

\parbox{0.4\textwidth}{\caption{\label{antiLambdaGlobalPolarization_eta}
{\small
(Color online) Global polarization of $\bar\Lambda$--hyperons as a function of $\bar\Lambda$ pseudorapidity.
Black circles show the results for Au+Au collisions 
at $\sqrt{s_{NN}}$=200~GeV (centrality region \mbox{20-70\%}) and 
red squares indicate the results for Au+Au collisions at 
$\sqrt{s_{NN}}$=62~GeV (centrality region \mbox{0-80\%}).
Only statistical errors are shown.
}}}
\end{center}
\end{figure}

\begin{figure}[h] 
\begin{center}
\includegraphics[width=0.5\textwidth]{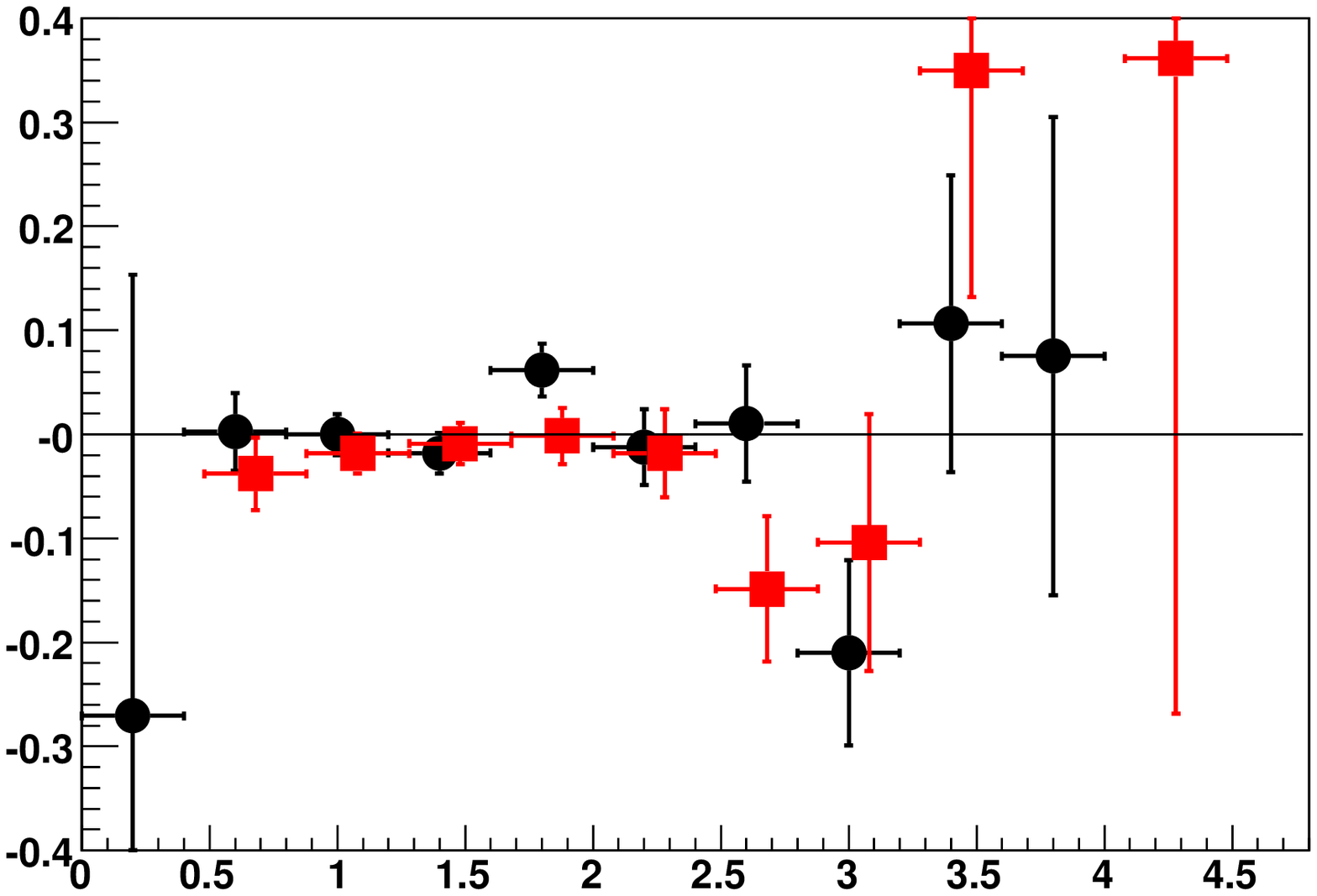} 
\put(-230,70){\rotatebox{90}{$P_{\bar\Lambda}$}} 
\put(-140,-5){$p_t$ (GeV/c)} 
\put(-195,125){{\bf STAR Preliminary}} 

\parbox{0.4\textwidth}{\caption{\label{antiLambdaGlobalPolarization_pt} 
{\small
(Color online) Global polarization of $\bar\Lambda$ hyperons as a function of $\bar\Lambda$ transverse momentum.
Black circles show the results for Au+Au collisions 
at $\sqrt{s_{NN}}$=200~GeV (centrality region \mbox{20-70\%}) and 
red squares indicate the results for Au+Au collisions at 
$\sqrt{s_{NN}}$=62~GeV (centrality region \mbox{0-80\%}).
Only statistical errors are shown.
}}}
\end{center}
\end{figure}

Figures \ref{antiLambdaGlobalPolarization_eta} and \ref{antiLambdaGlobalPolarization_pt} present
$\bar\Lambda$--hyperon's global polarization
as a function of $\bar\Lambda$ pseudorapidity and transverse momentum.
Black circles (red squares) show the result of the measurement for Au+Au collisions at 
$\sqrt{s_{NN}}$=200~GeV (62~GeV) with the STAR detector.
Within statistical errors no deviation from zero has been observed.
The $p_t$-integrated global polarization result is dominated by the region
$p^{\bar\Lambda}_t<3$~GeV, where measurements are consistent with zero.
The constant line fits for the $\bar\Lambda$--hyperon global
polarization as a function of pseudorapidity give: $P_{\bar\Lambda} = (1.7 \pm 10.7) \times 10^{-3}$ for Au+Au collisions
at $\sqrt{s_{NN}}$=200~GeV (centrality region \mbox{20-70\%}) and $P_{\bar\Lambda} = (-17.3 \pm
11.0) \times 10^{-3}$ for Au+Au collisions at $\sqrt{s_{NN}}$=62~GeV (centrality region
\mbox{0-80\%}).
These results are consistent with those from $\Lambda$--hyperon global polarization measurements~\cite{Selyuzhenkov:2006fc}.

\section{Acceptance corrections and systematic uncertainties}
\label{Systematic}
The derivation of Eq.~\ref{GlobalPolarizationObservable} assumes a perfect reconstruction acceptance for $\bar\Lambda$--hyperon.
For the case of non-perfect detector one has to correct results by detector acceptance function~\cite{Selyuzhenkov:2006tj}:
\begin{eqnarray}
\label{AccCoefficient}
A_{0}(p_t^H,\eta^H) = \frac{4}{\pi} \int {\frac{d\Omega^*_p}{4\pi} \frac{d\phi_H}{2\pi}A\left({\bf p}_H, {\bf p}^*_p\right) \sin\theta^*_p}.
\end{eqnarray}
Here $d\Omega^*_p = d\phi^*_p \sin \theta^*_p d
\theta^*_p$ is the solid angle of the hyperon decay products' 3-momentum ${\bf p}^*_p$ in the hyperon rest
frame; ${\bf p}_H$ ($\phi_H$) is the hyperon 3-momentum (azimuthal angle), and $A\left({\bf p}_H, {\bf p}^*_p\right)$ is
a function to account for detector acceptance.
For the $\bar\Lambda$--hyperons reconstructed with STAR detector, this function is found to follow the same
pseudorapidity and transverse momentum dependence as for $\Lambda$--hyperon~\cite{Selyuzhenkov:2006tj},
and corresponding corrections are estimated to be less than 20\%.
Similar to $\Lambda$--hyperon results~\cite{Selyuzhenkov:2006tj}, the admixture from
$\bar\Lambda$ directed flow to the global polarization measurement is found to be negligible.

Another type of correction is from possible dependence of the
hyperon global polarization on the relative azimuthal angle between the direction of hyperon's 3-momentum
and the system orbital momentum {\boldmath $L$}.
For the perfect detector the observable in Eq.~\ref{GlobalPolarizationObservable}
gives an average of the global polarization over this relative azimuthal angle.
This can be shown by expanding
the global polarization as a function of ($\phi_H-\Psi_{RP}$) in a sum
(due to the symmetry of the system only even harmonics contribute):
\begin{eqnarray}
\label{sumForGlobalPolarization}
&&P_H\left(\phi_H-\Psi_{RP},p_t^H,\eta^H\right)~=\\
\nonumber
&&=~\sum_{n=0}^\infty P_H^{(n)}\left(p_t^H,\eta^H\right)\cos\{2n[\phi_H-\Psi_{RP}]\}.
\end{eqnarray}
Global polarization averaged over all possible values of ($\phi_H-\Psi_{RP}$) will be given by:
\begin{eqnarray}
\label{GPaverage}
P_H\left(p_t^H,\eta^H\right) & \equiv & \overline{P_H\left(\phi_H-\Psi_{RP},p_t^H,\eta^H\right)}
\\ \nonumber & = & P_H^{(0)}\left(p_t^H,\eta^H\right).
\end{eqnarray}
For the case of an imperfect detector, the observable  in Eq.~\ref{GlobalPolarizationObservable}
will be proportional to $P_H^{(0)}$ and contains the additive admixture
of higher harmonic terms, namely $P_H^{(2)}$ (compare with Eq.~4 in~\cite{Selyuzhenkov:2006tj}):
\begin{eqnarray}
\label{GlobalPolarizationObservableAcc}
&& \langle \sin \left( \phi^*_p - \Psi_{RP}\right)\rangle =
\\ &&= \nonumber \frac{\alpha_H}{2} \int {\frac{d\Omega^*_p}{4\pi} \frac{d\phi_H}{2\pi}A\left({\bf p}_H, {\bf p}^*_p\right) \sin\theta^*_p}\times
\\ && \nonumber \times \left[P_H^{(0)} -\frac{P_H^{(2)}}{2} \cos\left[2(\phi_H-\phi_p^*)\right]\right].
\end{eqnarray}
For the perfect acceptance this leads to observable in Eq.~\ref{GlobalPolarizationObservable},
where under $P_H$ ($P_{\bar\Lambda}$) one have to understand $P_H^{(0)}$.
According to Eq.~\ref{GPaverage}, $P_H^{(0)}$ is
the average of global polarization over relative azimuthal
angle between hyperon's direction and the system's orbital momentum {\boldmath $L$}.

\begin{figure}[h]
\begin{center}
\includegraphics[width=0.5\textwidth]{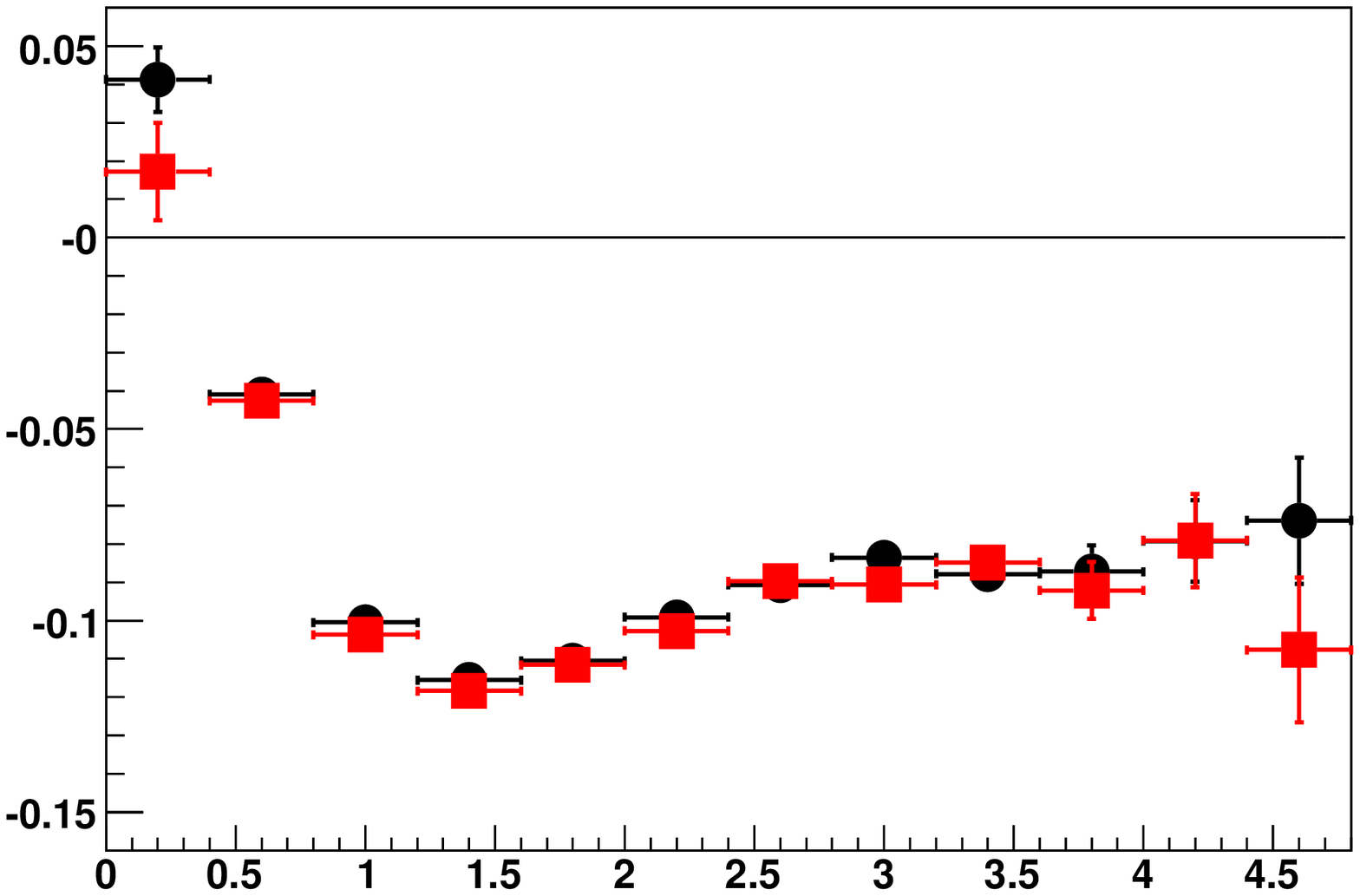}
\put(-233,60){\rotatebox{90}{$A_{2}^{\Lambda,\bar\Lambda}$}}
\put(-125,-5){$p_t^{\Lambda,\bar\Lambda}$}\\
\includegraphics[width=0.5\textwidth]{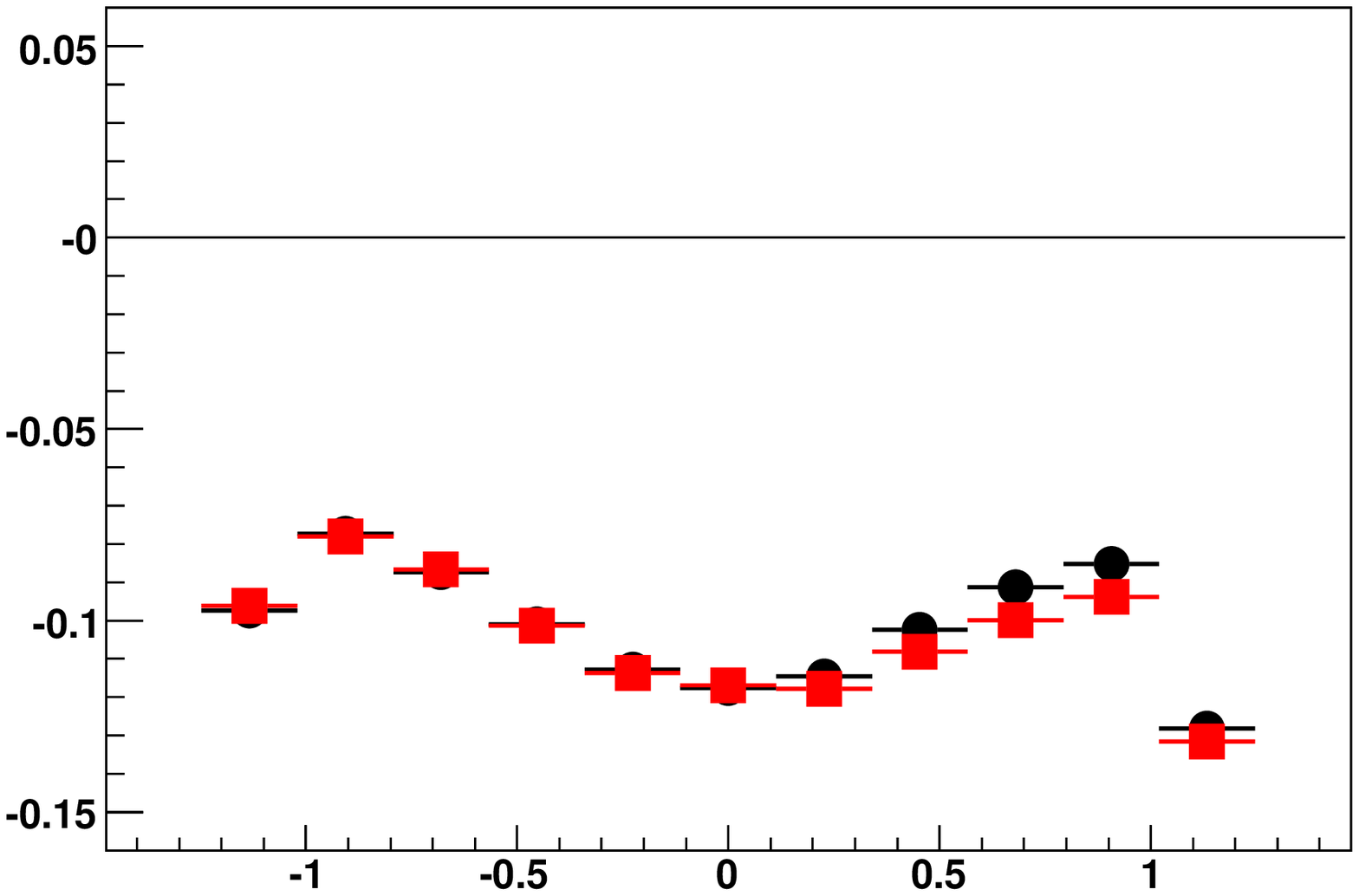}
\put(-233,60){\rotatebox{90}{$A_{2}^{\Lambda,\bar\Lambda}$}}
\put(-125,-5){$\eta^{\Lambda,\bar\Lambda}$}

\parbox{0.4\textwidth}{\caption{ \label{corr2Figure}
{\small
(Color online)
Integral (\ref{AccCoefficientAdditive}) as a function of $\Lambda$ (black circles) and
$\bar\Lambda$ (red squares) transverse momentum (top) and pseudorapidity (bottom).
}}}
\end{center}
\end{figure}

Due to the non-uniform detector acceptance, Eq.~\ref{GlobalPolarizationObservableAcc} contains two different contributions.
First one is defined by acceptance correction function $A_0(p_t^H,\eta^H)$ in Eq.~\ref{AccCoefficient}.
Deviation of this function from unity (perfect detector) affects the overall scale of the measured global polarization.
The contribution from the second term is proportional to $P_H^{(2)}$ and defined by the function:
\begin{eqnarray}
\label{AccCoefficientAdditive}
A_{2}(p_t^H,\eta^H) &=& \frac{2}{\pi} \int {\frac{d\Omega^*_p}{4\pi} \frac{d\phi_H}{2\pi}A\left({\bf p}_H, {\bf p}^*_p\right)}\times
\\&&\nonumber  \times \sin\theta^*_p \cos\left[2(\phi_H-\phi_p^*)\right].
\end{eqnarray}
For perfect acceptance this function is zero.
Taking into account that the background contribution to the $\Lambda$ and $\bar\Lambda$
invariant mass distribution is less than 8\%,
the value of function $A_2(p_t^H,\eta^H)$ can be extracted
directly from the experimental data by calculating
$\left<\sin\theta^*_p\cos\left[2(\phi_H-\phi_p^*)\right]\right>$ for $\Lambda$ and $\bar\Lambda$ candidates.
The result of such calculations is presented in Figure~\ref{corr2Figure}.
Assuming that different terms in expansion (\ref{sumForGlobalPolarization}) are of the same order of magnitude,
the corresponding corrections from the admixture of $P_H^{(2)}\left(p_t^H,\eta^H\right)$ to
the $\Lambda$ and $\bar\Lambda$'s hyperon global polarization measurement
are found to be less than 20\%.

Feed-down effects, spin precession, uncertainty of reaction plane angle reconstruction procedure
for the $\Lambda$--hyperon global polarization measurement
has been discussed in~\cite{Selyuzhenkov:2006tj}.
Based on the similar study, the overall relative uncertainty in the $\bar\Lambda$ global polarization measurement
due to detector effects is found to be less than a factor of 2.

\section{Conclusion}
\label{Conclusion} 
The $\bar\Lambda$--hyperon global polarization has been measured in Au+Au
collisions at center of mass energies $\sqrt{s_{NN}}$=62 and 200~GeV with the STAR detector at RHIC.
Within uncertainties we observe no significant deviation from zero of the $\bar\Lambda$ global polarization.
The possible dependence of the global polarization on relative azimuthal angle
between system's orbital momentum and hyperon's 3-momentum is discussed.
The corresponding systematic uncertainty due to detector acceptance
in the $\Lambda$ and $\bar\Lambda$ global polarization measurements
 is found to be less than 20\%.

Combining results of this measurement and those from~\cite{Selyuzhenkov:2006fc}, 
an upper limit of $|P_{\Lambda,\bar\Lambda}| \leq 0.02$ for the global
polarization of $\Lambda$ and $\bar\Lambda$ hyperons within STAR's acceptance is 
obtained. The obtained upper limit is far below the few tens of percent values discussed 
in~\cite{LiangPRL94}, but it falls within the predicted region from the more realistic 
calculations~\cite{Liang:Xian2006} based on the HTL (Hard Thermal Loop) model.

\end{document}